\documentclass[twocolumn]{aastex63}
\usepackage{autobreak}
\usepackage{breqn}

\newcommand{\nn}{\nonumber}
\newcommand{\letter}{paper}
\newcommand{\be}{\begin{equation}}
\newcommand{\ee}{\end{equation}}
\newcommand{\spec}{{\rm spec}}
\newcommand{\zm}{z_{\rm m}}

\received{\today}
\shortauthors{Tonegawa \& Okumura}
\begin{document}

\title{
First Evidence of Intrinsic Alignments of Red Galaxies at $z > 1$:\\
Cross-correlation between CFHTLenS and FastSound Samples
}

\author{Motonari Tonegawa}
\affiliation{Asia Pacific Center for Theoretical Physics, Pohang, 37673, Korea}
\author{Teppei Okumura}
\affiliation{Institute of Astronomy and Astrophysics, Academia Sinica, No. 1, Section 4, Roosevelt Road, Taipei 10617, Taiwan}
\affiliation{Kavli Institute for the Physics and Mathematics of the Universe (WPI), UTIAS, The University of Tokyo, Kashiwa, Chiba 277-8583, Japan}



\begin{abstract}
We report the first evidence for intrinsic alignment (IA) of red galaxies at $z>1$. 
We measure the gravitational shear-intrinsic ellipticity (GI) cross-correlation function at $z\sim1.3$ using galaxy positions from the FastSound spectroscopic survey and galaxy shapes from Canada-Hawaii-France telescope lensing survey data.
Adopting the non-linear alignment model, we obtain a $2.4\sigma$-level detection of the IA amplitude $A^{\rm LA}=27.48_{-11.54}^{+11.53}$ (and $2.6\sigma$ with $A^{\rm LA}=29.43_{-11.49}^{+11.48}$ when weak-lensing contaminations are taken into account),
which is larger than the value extrapolated from the constraints obtained at lower redshifts.
Our measured IA is translated into a $\sim 20\%$ contamination to the weak lensing power spectrum for the red galaxies.
This marginal detection of IA for red galaxies at $z>1$ motivates
the continuing investigation of the nature of IA for weak lensing studies.
Furthermore, our result provides the first step to utilize IA measurements in future high-$z$ surveys as a cosmological probe, complementary to galaxy clustering and lensing. 

\end{abstract}



\section{Introduction} \label{sec:intro}
Intrinsic alignment (IA) is a coherent alignment of galaxy orientations with the surrounding large-scale structure caused by the local gravitational interaction \citep{Croft:2000,Heavens:2000,Catelan:2001,Hirata:2004}.
IA has been considered as one of the main contaminants of weak lensing surveys, where the source galaxies
are assumed to be randomly oriented.
While weak lensing is a major probe to constrain cosmological parameters, 
it requires an accurate modeling of IA to avoid constraints being biased \citep{
Joachimi:2015},
and thus we need to understand how large the IA effect is and how it depends on scales and time.
IA also contains useful information about galaxy formation and evolution.
Galaxies at higher redshifts could be either more strongly or weakly aligned because of shorter time elapsed for internal and external interactions, e.g., structure formations and mergers, to boost or suppress IAs.
Therefore, measuring the IA at various redshifts will help constrain dynamical aspects of galaxy evolution models.

Furthermore, due to the fact that galaxy shapes are linearly related the gravitational potential on cosmological scales, there is a growing interest of using IA as a new tool to constrain cosmological models \citep{
%Schmidt:2012,
Chisari:2013,
%Okumura:2019,
Taruya:2020}.
Since ongoing and future deep surveys provide high-quality galaxy images toward high redshifts, %\citep[e.g.,][]{Aihara:2018}, 
IA can be a powerful cosmological probe complementary to galaxy clustering and weak lensing. 

The IA of elliptical galaxies has been observed at $z<1$
by measuring intrinsic ellipticity correlations
and shown to contaminate the weak-lensing power spectrum by $\sim 10\%$ \citep{Mandelbaum:2006,Hirata:2007,Okumura:2009,Okumura:2009a,Singh:2015,Johnston:2021}.
Extending such observations to higher redshifts is important to better understand IA as a cosmological probe and as a source of contamination to weak lensing studies.

Motivated by these, in this \letter, we report the first
possible evidence for IA of red galaxies at $z>1$ by cross-correlating the shape sample selected from the Canada-France-Hawaii Telescope Lensing Survey (CFHTLenS; \cite{Erben:2013}) data with the galaxy density sample from the Subaru FastSound survey \citep{Tonegawa:2015}.

The cosmological parameters used in this paper are $(\Omega_m,\Omega_\Lambda,h,\sigma_8)=(0.3,0.7,0.7,0.8)$,
except when we use $h=1.0$ to calculate the absolute luminosity for consistency with preceding studies.
All the distances and separations are expressed in comoving units.

\section{Data}\label{section:data}
\subsection{CFHTLenS shape sample}\label{section:ellip}

Current spectroscopic samples of elliptical galaxies for high redshifts are not sufficiently large because one needs a long exposure time to observe the $4000{\rm \AA}$ breaks.  
We thus use the publicly available
CFHTLenS \citep{Erben:2013} data
for the shape sample.\footnote{http://www.cadc-ccda.hia-iha.nrc-cnrc.gc.ca/en/community/CFHTLens/query.html}
It provides accurate photometric redshifts (photo-$z$) \citep{Hildebrandt:2012} and shape measurements for galaxies covering $154~\rm{deg}^2$ over the CFHT Wide fields.
The typical photo-$z$ scatter is $\sigma_{z}/(1+z)\sim0.04$ at $z>1$.
We take the data in the photo-$z$ range, $1.13<z_{\rm ph}<1.63$, which 
covers the entire redshift range of
the spectroscopic FastSound sample described below.
We also limit the data to the angular regions overlapping with the FastSound sky coverage, explained in the following subsection.
We exclude stars from our sample, which are assigned zero ellipticity.
Galaxies in the catalog are classified by the spectral type, symbolized as a numeric flag \texttt{T_B},\footnote{For the correspondence between \texttt{T_B} and galaxy types, see \citet{Erben:2013}} estimated by the Bayesian photometric redshift code \citep[BPZ:][]{Benitez:2000}
using the template set of \citet{Capak:2004}.
Smaller (larger) \texttt{T_B}
values correspond to redder (bluer) galaxies, and
we adopt $\texttt{T_B}<1.5$ to select elliptical galaxies which are expected to have strong IA signals.
The criterion results in $11320$ galaxies, $12\%$ of all the galaxies in the redshift range.

The two-component ellipticity of galaxies is given by $(e_+, e_\times) =\frac{1-q}{1+q}(\cos{2\phi},\sin{2\phi})$, where $q$ is the minor-to-major-axis ratio on the celestial sphere and $\phi$ is the position angle. 
The ellipticity is 
estimated using {\it lensfit} \citep{Miller:2007} applied to the $i'$-band image and corrected for the additive and the multiplicative biases as done in \citet{Tonegawa:2018}.

\subsection{FastSound spectroscopic sample}\label{section:fastsound}
The FastSound is a spectroscopic survey with the FMOS instrument \citep{Iwamuro:2012} to measure redshifts of $\sim 4000$ blue star-forming galaxies at $z\sim1.36$ \citep{Tonegawa:2015,Okada:2016}
in $\sim 25~{\rm deg}^2$ of the CFHT Wide fields.
Out of the four regions of the CFHT Wide, we use the W2 and W3 fields because these fields cover the majority of the FastSound.
As the survey used near-infrared spectroscopy targeting H$\alpha$6563, the galaxy sample ranges from $z=1.19$ to $1.55$.
We select galaxies that have emission line features with signal-to-noise ratio greater than $4.5$, obtaining $2665$ objects.
To measure correlation functions we use the random catalog constructed in \citet{Okumura:2016}.

\section{Galaxy-intrinsic ellipticity correlation function} \label{section:measurement}
This study focuses on the galaxy-intrinsic ellipticity (GI) correlation, 
with the density field from the FastSound galaxy sample and the ellipticity field from the CFHTLenS photo-$z$ galaxy sample. 
We do not consider the auto-correlation of the intrinsic ellipticity (II) because it is noisier than the GI correlation and more severely affected by the photo-$z$ uncertainties in our analysis.  

We propose an estimator for the GI cross-correlation function, 
which is an extension of the Hamilton estimator of the galaxy cross-correlation function \citep{Hamilton:1993a,Wang:2011}:
\begin{equation}\label{equation:estimator_H93}
\xi_{{\rm g}+}(\bar{r}_p,\bar{r}_\pi) = \frac{S^{+}Q\cdot RR-S^{+}R\cdot QR}{QR\cdot DR},
\end{equation}
where $(\bar{r}_p,\bar{r}_\pi)$ are the transverse and parallel separations of galaxy pairs, respectively.
The term $RR$ is the pair count of the random sample,
$QR$ ($DR$) is the cross pair count between spectroscopic (photometric) and random samples, 
and $S^{+}Q$ ($S^{+}R$) denotes the sum of the tangential shear component, $e_+$, 
redefined relative to the direction to the spectroscopic (random) sample. 
To account for the uncertainty of shape measurements, $e_+$ is weighted by the inverse-variance.
We put a bar on the separations to explicitly denote that they are affected by photo-$z$ (see section \ref{section:la_model} for their relation to spectroscopically determined $(r_p, r_\pi)$).

We integrate the 3-D GI correlation function along the line of sight to minimize the effect of both photo-$z$ errors and redshift space distortions, 
\begin{equation}\label{equation:GI_proj}
w_{\rm g+}(\bar{r}_p)=\int_{-\bar{r}_{\pi,{\rm max}}}^{\bar{r}_{\pi,{\rm max}}} \xi_{\rm g+}(\bar{r}_p,\bar{r}_\pi)d\bar{r}_\pi,
\end{equation}
where the value of $\bar{r}_{\pi, {\rm max}}$ needs to be chosen to include %all
the correlated pairs scattered by the photo-$z$ errors\footnote{This effect elongates the correlation function along the line of sight, similarly to the small-scale redshift-space distortion, where virialized galaxy motions cause random displacements of observed galaxy positions \citep{Hamilton:1998}.}
while minimizing large-scale noise.
We carefully tested the integral range and found that $\bar{r}_{\pi,{\rm max}}=160\ h^{-1}{\rm Mpc}$ with the linear bin width $\Delta \bar{r}_\pi = 5\ h^{-1}{\rm Mpc}$ provides the highest $S/N$ ratio.
We will discuss the choice of $\bar{r}_{\pi,{\rm max}}$ more in section \ref{section:results}.
We also measured
$w_{g\times}$ by replacing $e_+$ with $e_\times$ and use it for systematic tests because this quantity
should be zero at all scales.
We have also tried another estimator for the correlation function \citep{Mandelbaum:2006,Joachimi:2011},
\begin{equation}\label{equation:estimator_M06}
\xi_{\rm g+}(\bar{r}_p,\bar{r}_\pi) = \frac{S^{+}Q-S^{+}R}{DR}.
\end{equation}
We have confirmed that the two estimators gave the same results (see appendix \ref{section:appendix_systematics}).

The covariance matrix for the GI correlation function is estimated using 
82 jackknifed realizations as done in \citet{Tonegawa:2018}.
The survey regions were split into $33$ ($49$) sub-regions for W2 (W3){, each with a side length of $\sim 25 h^{-1}{\rm Mpc}$.
The covariance matrix is estimated for each of the W2 and W3 realizations 
and they are combined following the inverse-variance weighting \citep[see e.g.,][]{Okumura:2021}.

In the FastSound survey, some targets are not assigned fibers due to their finite number and
this affects the correlation function measurement at an angular scale of $\sim 1'$ \citep{Okumura:2016}. However, this effect is significantly alleviated by considering the cross-correlation with a photo-$z$ sample.

\section{Results and Discussions}\label{section:results}
\subsection{The GI Correlation Functions}\label{section:correlations}
\begin{figure}
 \begin{center}
  \includegraphics[width=0.45\textwidth]{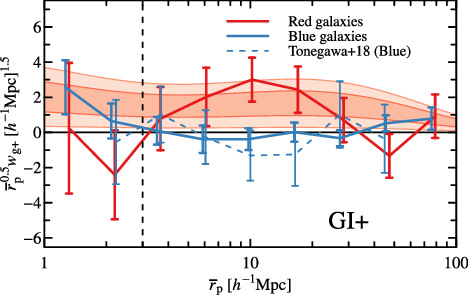}
  \vspace{11pt}\\
  \includegraphics[width=0.45\textwidth]{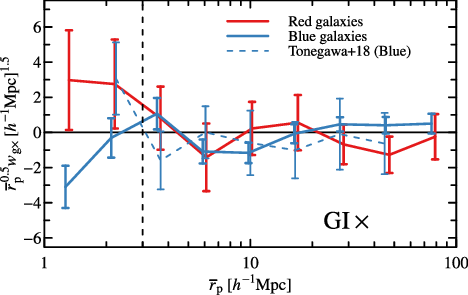}
 \end{center}
 \caption{Projected correlation functions, $w_{g+}$ (top) and $w_{g\times}$ (bottom) as a function of transverse separation $\bar{r}_p$.
We use the FastSound sample for the galaxy density field for all the measurements presented here. 
We use the different shape samples as the galaxy shape field for different lines: 
red galaxies (red lines) and blue galaxies (blue solid lines).
The blue dashed line is the GI correlation of the FastSound blue galaxies measured by \citet{Tonegawa:2018}.
The dark and light red shaded regions in the top panel indicate the $68\%$ and $95\%$ confidence intervals of the best-fitting NLA model obtained for red galaxies at $\bar{r}_p> 3~h^{-1}{\rm Mpc}$ denoted by the vertical lines.
The error bars are obtained from the diagonal elements of the covariance matrix, $C_{ii}^{1/2}$.
}
\label{figure:GI}
\end{figure}

We show the GI cross-correlation function 
between the FastSound galaxy positions and CFHT red galaxy shapes,
$w_{g+}$, as the solid red line in the upper panel of figure \ref{figure:GI}. 
We take nine logarithmic bins from $\bar{r}_p=1$ to $100~h^{-1}{\rm Mpc}$. 
The error bars represent $1-\sigma$ uncertainties estimated from the 82 jackknifed realizations. 
We see a small but non-zero signal in the $w_{g+}$ measurement.
The positive correlation means that the major axes of
CFHT red galaxies
tend to point toward overdensities. 
The bottom panel of figure \ref{figure:GI} shows 
the other GI cross-correlation function, $w_{g\times}$, which should vanish on all scales.
The signal of $w_{g\times}$ is consistent with zero beyond $3~h^{-1}{\rm Mpc}$. 
We perform a further test by shuffling the ellipticity information in the shape sample before measuring the GI correlation and find that the resulting correlation is indeed consistent with zero beyond the scale, as shown in Appendix \ref{section:appendix_systematics}.
We thus consider our measurements free from observational systematics at $\bar{r}_p>3~h^{-1}{\rm Mpc}$ and use this scale for the following analysis. 

To see the detection significance for the IA signal, we fit a power-law model, $w_{g+}^{\rm PL}(\bar{r}_p) = (1-f_{\rm blund})A^{\rm PL} \left (\frac{\bar{r}_p}{20~h^{-1}{\rm Mpc}} \right )^{\gamma}$, to the measured correlation function.
The parameter $f_{\rm blund}$ is a fraction of redshift blunders (noise lines and ${\rm O_{III}}$ doulets) and $f_{\rm blund}=0.071$ for
our FastSound sample \citep{Okada:2016,Okumura:2016}.
Taking account of the full covariance matrix,
we calculate $\chi^2$ statistics in the range of $3 < \bar{r}_p < 100~h^{-1}{\rm Mpc}$, where $w_{g\times}$ 
is consistent with zero.
We fix $\gamma=-0.88$, as obtained by \citet{Hirata:2007} for
luminous red galaxies (LRGs) at $z\simeq 0.3$.
This is a reasonable assumption because we obtain $\gamma = -0.75^{+0.42}_{-0.42}$
when we simultaneously determine $A^{\rm PL}$ and $\gamma$.
The resulting constraint on the amplitude parameter is
$A^{\rm PL}=0.266_{-0.116}^{+0.112}$ (68\% CL), corresponding to a $2.3 \sigma$ detection of IA.

Our finding is robust because a cross-correlation function tends to be uncorrelated between two independent samples and thus less sensitive to systematic effects.
Furthermore, we have varied several parameters 
to confirm that the signal is still detectable. Specifically, we have applied $\bar{r}_{\pi, {\rm max}}=200~h^{-1}{\rm Mpc}$ and confirmed the signal at $\sim2\sigma$.
Also, changing the $\texttt{T_B}$ threshold to $1.1$ (selecting a redder population) resulted in a similar detection significance.
Therefore, we conclude that the signal indeed indicates the evidence of IA.

Unlike red galaxies, we do not find non-zero GI correlations for the CFHT blue galaxy shapes, selected by the criteria of $2.0<\texttt{T_B}<4.0$, as seen as the solid blue line in figure \ref{figure:GI}.
This is consistent with the result of \citet{Tonegawa:2018}, as shown as the blue dashed line for comparison, who measured the GI auto-correlations of blue galaxies that are spectroscopically confirmed from the FastSound at $z\sim 1.36$. Other studies at lower redshifts also have not found any GI signal for blue galaxies \citep{Mandelbaum:2011, Johnston:2021}.
IA of spiral galaxies are likely to be explained by the quadratic alignment model \citep{Catelan:2001, Hirata:2004, Hirata:2010, Kirk:2015},
and the model indeed predicts null GI signals for a Gaussian density field.

\subsection{Linear Alignment Model}\label{section:la_model}
Here we consider a more physically motivated prediction of IA, the linear alignment (LA) model  \citep{Catelan:2001,Hirata:2004} 
which relates the shear field linearly to the gravitational potential. Under this model the density- intrinsic ellipticity power spectrum at redshift $z$ is given by 
\begin{equation}\label{equation:LA}
P_{\delta {\rm I}}(k,z)=\frac{C_1 \bar{\rho}(z)}{(1+z)D(z)}a^2P_\delta(k,z),
\end{equation}
where $\bar{\rho}(z)$ is the mean matter density, $D(z)$ is the growth factor, and $P_\delta(k,z)$ is the matter power spectrum.
While the original LA model used linear theory prediction for $P_\delta$ \citep{Hirata:2004}, using the non-linear matter power spectrum was found to better explain the observed IA \citep[non-linear LA, NLA;][]{Bridle:2007,Blazek:2011}. Therefore, we use the non-linear matter spectrum of \citet{Takahashi:2012} to obtain the theoretical prediction.
The normalization parameter $C_1$ varies much with given galaxy samples. Following the convention, we introduce another parameter, $A^{\rm LA}$, as $A^{\rm LA}=C_1\rho_{\rm cr}/0.0134$, where $\rho_{\rm cr}$ is the critical density.

The Hankel transform converts the power spectrum into the 3-D gI correlation function \citep{Okumura:2020,Okumura:2020a}:
\begin{equation}\label{equation:gI_spec}
\xi_{\rm gI}^{\spec}(r_p,r_\pi,z)=(1-\mu^2) b_g\int_{0}^{\infty} \frac{k^2dk}{2\pi^2} P_{\delta {\rm I}}(k,z)j_2(kr),
\end{equation}
where $\mu=r_\pi/r$ with $r=\sqrt{r_p^2+r_\pi^2}$, $j_2$ is the spherical Bessel function of the second order, and $b_g$ is the linear bias parameter of the FastSound galaxies, $b_g=1.9$ \citep{Okumura:2016}.
We use photo-$z$ for the shape sample, 
which modulates equation~(\ref{equation:gI_spec}) due to the scatter along the line-of-sight as \citep{Joachimi:2011}
\begin{align}\label{equation:xig+bar}
\xi_{\rm gI}
 (\bar{r}_p,\bar{r}_\pi,& \bar{z}_{\rm m}) = % \nn \\
 \int dz_2 \;  
p_\epsilon \left(z_2|\bar{z}_2\right) \nn \\
&\times\; \xi_{\rm gI}^{\spec} \left(\bar{r}_p\; \frac{\chi (\zm)}{\chi (\bar{z}_{\rm m})},\frac{c\; |z_2-z_1|}{H(\zm)} ,\zm\right)\;,
\end{align}
where $\bar{z}_{\rm m}$ denotes the mean of photo-$z$ of the shape sample and spec-$z$ of the density sample, a bar means a quantity affected by photo-$z$,
$H$ is the Hubble parameter,
$c$ is the speed of light, $\chi(z)$ is the comoving distance, $z_1=\bar{z}_{\rm m}-\bar{r}_{\pi} H(\bar{z}_{\rm m})/2c$, and $p_\epsilon(z|\bar{z})$ denotes the probability distribution of the true redshift $z$
for a given photo-$z$, $\bar{z}$, for the shape sample.
We assume that the error in photo-$z$ follows the normal distribution with $\sigma_z/(1+\bar{z})=0.04$ \citep{Hildebrandt:2012}.
We integrate $\xi_{\rm gI}
(\bar{r}_p,\bar{r}_\pi,\bar{z}_{\rm m})$ along the line of sight to obtain the projected correlation function $w_{g+}(\bar{r}_p)
$, similarly to equation (\ref{equation:GI_proj}).
With our choice of ${\bar{r}_{\pi,{\rm max}}}$, ${\bar{r}_{\pi,{\rm max}}} = 160~h^{-1}{\rm Mpc}$, we find the amplitude of $w_{g+}(\bar{r}_p)$ becomes $76\%$ of that determined with spec-$z$, $w^\spec_{g+}(r_p)$. 
As shown by \citet{Joachimi:2011}, the $r_p$ dependence remains almost unchanged
when photo-$z$ are considered.
The LA model fitting to the measured $w_{g+}(\bar{r}_p)$ gives a constraint on the amplitude as $A^{\rm LA} =27.48_{-11.54}^{+11.53}$, showing a $2.4\sigma$ deviation from zero similarly to the result obtained in section \ref{section:correlations}. 
The dark and light red shaded regions in the top panel of figure \ref{figure:GI} indicate the $68\%$ and $95\%$ confidence levels of the NLA model.

The observed galaxy shape is the sum of the intrinsic shape and the weak lensing shear, and the galaxy density is the sum of the intrinsic one and the lensing magnification effect. Thus, not only the gI signal but also galaxy-galaxy lensing (gG), magnification-shear correlations (mG), and magnification-intrinsic correlations (mI) contribute to the observed galaxy-shape correlation (see equation~(\ref{eq:angular_ne})).
Following \citet{Joachimi:2011}, we calculate the contributions of the gG, mG, and mI correlations taking into account photo-$z$ errors on the shape sample, as summarized in Appendix \ref{sec:limber}.
We obtain $\alpha_s=2.56$ for equation~(\ref{eq:limber_2d}) using our FastSound density sample at the $z$-band magnitude of $\sim23$, corresponding to the magnitude limit of the sample.
Including all the lensing effects, our constraint on $A^{\rm LA}$ becomes $A^{\rm LA}=29.43_{-11.49}^{+11.48}$, with each contribution being $1.3\%$ (gG), $5.8\%$ (mG), and $0.5\%$ (mI). 
Since our measurement is at a relatively high redshift, $z\sim 1.3$, the mG term becomes the dominant contamination.
Since there is an uncertainty in determining the faint-end slope $\alpha_s$,
we quote the constraint without considering the lensing effect, $A^{\rm LA}=27.48^{+11.53}_{-11.54}$, as our main result. This corresponds to a conservative constraint because the sign of the gG and mG terms is opposite of the gI term.

\begin{table*}
\footnotesize
\begin{center}
  \caption{Summary of the IA amplitude for various sample redshift and luminosity.
The mean absolute $r$-band magnitudes (with $h=1.0$) are $k$ and $e$-corrected.
The fourth column presents the amplitude of the LA model of equation (\ref{equation:LA}), whereas the fifth column presents the amplitude parameter in terms of the LA model in \cite{Hirata:2004} (HS04) to facilitate comparisons with previous works. Note that the values in these columns are proportional with the factor of $(1+\langle z \rangle)^2$.}
\begin{tabular}{cccccc}
\hline
\hline
 Data & $\langle z \rangle$ & $\langle M_r \rangle$ & $A^{\rm LA}$ & $A^{\rm LA}$ (HS04) & Reference \\
\hline
SDSS main, red & $0.12$ & $-19.88$ & $2.50^{+0.77}_{-0.73}$ & $1.99^{+0.61}_{-0.58}$ & \citet{Johnston:2019}\\
GAMA $z<0.26$, red & $0.17$ & $-20.47$ & $3.63^{+0.79}_{-0.79}$ & $2.65^{+0.58}_{-0.58}$ & \citet{Johnston:2019}\\
BOSS LOWZ L1 & $0.28$ & $-21.70$ & $8.5^{+0.9}_{-0.9}$  & $5.2^{+0.5}_{-0.5}$ & \citet{Singh:2015} \\
BOSS LOWZ L2 & $0.28$ & $-21.27$ & $5.0^{+1.0}_{-1.0}$  & $3.1^{+0.6}_{-0.6}$ & \citet{Singh:2015} \\
BOSS LOWZ L3 & $0.28$ & $-21.07$  & $4.7^{+1.0}_{-1.0}$  & $2.9^{+0.6}_{-0.6}$  & \citet{Singh:2015} \\
BOSS LOWZ L4 & $0.28$ & $-20.76$  & $2.2^{+0.9}_{-0.9}$  & $1.3^{+0.5}_{-0.5}$ & \citet{Singh:2015} \\
GAMA $z>0.26$, red & $0.33$ & $-21.64$ & $3.55^{+0.90}_{-0.82}$ & $2.01^{+0.51}_{-0.46}$ & \citet{Johnston:2019}\\
MegaZ-LRG & $0.54$ & $-21.96$ & $4.51^{+0.64}_{-0.63}$  & $1.98^{+0.28}_{-0.28}$ & \citet{Joachimi:2011} \\
CFHT and FastSound & $1.31$ & $-20.54$ & $27.48_{-11.54}^{+11.53}$ &  $5.15^{+2.16}_{-2.16}$ & This work \\
\hline
\end{tabular}
\label{table:NLA}
\end{center}
\end{table*}

\begin{figure}
 \begin{center}
  \includegraphics[width=0.45\textwidth]{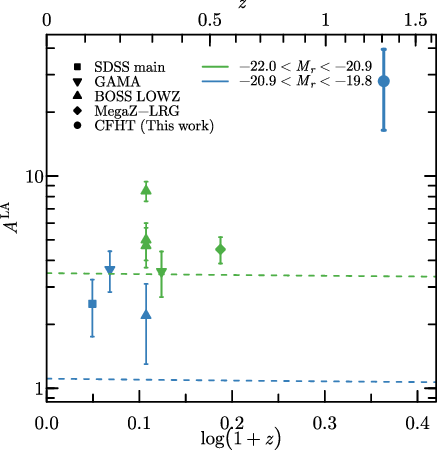}
 \end{center}
 \caption{Constraints on the amplitude of the NLA model as a function of redshift. 
 The points are color-coded according to the mean sample luminosity, and different symbols are assigned to different survey samples (see Table \ref{table:NLA}).
 The dashed lines are the best-fitting model prediction at $z=0.54$ obtained by \citet{Joachimi:2011} 
 (equation (\ref{equation:A_LA_pl})) for 
 $\langle M_r \rangle=-21.45$ (green) and $-20.35$ (blue).}
\label{figure:A_LA}
\end{figure}

\subsection{IA of
red galaxies as a function of redshift}
Table \ref{table:NLA} and figure \ref{figure:A_LA} show the constraints on $A^{\rm LA}$ 
obtained from our analysis at $z\sim 1.3$ together with the previous studies at lower redshifts at $z<1$. 
Since the amplitude of IA of galaxies strongly depends on the luminosity, 
we apply Galactic extinction \citep{Schlegel:1998} and $k+e$ corrections
to the mean $r$-band absolute magnitude, $\langle M_r \rangle$,
for a fair comparison of the IA of
red galaxies at different redshifts.
For the $k$-correction, we interpolate the templates between ellipticals ($\texttt{T_B}=1.0$) and Sbc galaxies ($\texttt{T_B}=2.0$) from \citet{Coleman:1980} to create the spectral energy density (SED) template of $\texttt{T_B}=1.091$, which represents our shape sample.
For the 
$e$-correction, 
we use the \textsc{Pegase} code \citep{Fioc:1999} to track the evolution of the $r'$-band magnitude of
elliptical galaxies by stellar population synthesis modeling.
We assume an instantaneous starburst model with the initial mass function of \citet{Scalo:1986}, solar metallicity, and the galaxy age of $12$ Gyr at $z=0$.
After $k+e$ corrections are made, $\langle M_r \rangle$ is modified from $-18.81$ to $-20.54$.

\citet{Joachimi:2011} examined the luminosity and redshift dependences of $A^{\rm LA}$ using a parametric form,
\begin{equation}\label{equation:A_LA_pl}
A^{\rm LA}(L_r,z)=\alpha \left (\frac{L_r}{L_r^p} \right)^\beta \left(\frac{1+z}{1+z_0} \right)^\eta,
\end{equation}
where $z_0=0.3$ and the $r$-band pivot luminosity $L_r^p$ is set to the value which corresponds to the absolute magnitude of $M_r=-22$.
They obtained $\alpha=5.76^{+0.60}_{-0.62}$, $\beta=1.13^{+0.25}_{-0.20}$, and $\eta=-0.27^{+0.80}_{-0.79}$ from the measurements of elliptical galaxies up to $z\sim0.5$. \citet{Singh:2015} reached a similar result using the BOSS-LOWZ data.
Since the constraint on $\eta$ in their study was not strong due to the limited redshift range probed, 
the scenario that the IA amplitude increases toward higher redshifts is still allowed,
as suggested by recent simulations \citep{Chisari:2016a, Samuroff:2021} and observations \citep{Yao:2020}.
Nevertheless, our constraint on $A^{\rm LA}$ at $z\sim 1.3$ is marginally larger than the prediction made at lower redshifts by \citet{Joachimi:2011}, by $\sim 2\sigma$ level.

There are several possible explanations for this discrepancy though it is not significant given the relatively large error bars of our measurement.
First, 
%there could be a selection bias.
Since the probed redshift is high,
faint galaxies may fall below the survey limit,
making the sample biased toward bright galaxies,
which are known to have higher IA amplitudes \citep{Singh:2015}.
Second, the $r'$-band magnitude may not be proper to represent the shape sample of high-redshift galaxies.
The $r'$-band magnitude is known as a good proxy of the stellar mass \citep{Mahajan:2018, Du:2020}, but it corresponds to $\lambda \sim 2000$--$3000 \AA$ at $z\sim1.3$, which is below the $4000 \AA$ break.
We only expect weak spectral energy density, regardless of the actual stellar mass, and hence $r'$-band magnitudes may not represent the stellar and total mass.
The $k-$correction accounts for the wavelength dependence,
but it will suffer from a large uncertainty due to the uncertainty of the depth of the SED breaks.
For a consistent comparison of the IA strength over a wide redshift range,
other physical quantities such as the stellar mass and galaxy bias may be preferable.
Of course, the redshift dependence of $A^{\rm LA}$ could be purely physical;
galaxy/halo interactions such as mergers can reduce IAs toward $z=0$, as seen in $N$-body simulations
\citep{Kurita:2021}.
Obtaining data points at different redshifts (e.g., $z\sim 0.8$) is desirable to explore this possibility.

\subsection{Contaminantion to weak lensing measurements}\label{section:contamination}

Finally, we examine how the measured GI at $z\sim 1.3$ could contaminate weak lensing signals.
Figure \ref{figure:GG_GI} presents the comparison between the expected weak lensing power spectrum ($C_{\rm GG}$) and two predictions of contamination from the IA ($C_{\rm GI}$), one being the $2\sigma$ upper limit ($A^{\rm LA}\sim50$) and the other the best-fitting model ($A^{\rm LA}\sim 25$).
We use the redshift distribution of the CFHTLenS sample to compute the angular power spectra. 
The contamination estimated from the best-fitting LA model is $C_{\rm GI}/C_{\rm GG}=23\%$ ($26\%$) at $l=100$ ($5000$).
We follow the procedure carried out by \cite{Tonegawa:2018} to infer the resulting systematic error on $\Omega_m$ and $\sigma_8$ and find $\Delta \sigma_8=-0.076$ and $\Delta \Omega_m=-0.054$
for $A^{\rm LA}=27.48$ when $100<l<5000$ is used.
At $z>1$, the fraction of late-type galaxies that have much weaker IA increases,
and thus the actual contamination will be smaller. Nevertheless, since upcoming lensing surveys aim at sub-percent precision, the effect of IA should be taken into account properly.

\begin{figure}
 \begin{center}
  \includegraphics[width=0.45\textwidth]{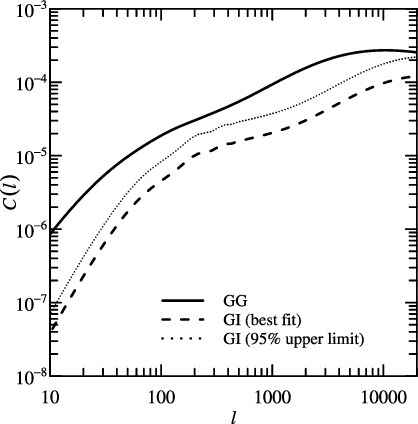}
 \end{center}
 \caption{Forecast of the angular power spectrum of GG (solid) and GI (lower dashed) terms. 
 The redshift distribution of the shape sample is used to compute them.
 The GI power spectrum is based on the best-fitting NLA model. 
The maximum contribution of the GI spectrum within the 95\% confidence level is 
shown as the dotted line. 
}
\label{figure:GG_GI}
\end{figure}

\section{Conclusions}\label{section:conclusions}
We have reported the first possible evidence for IA of elliptical galaxies at $z>1$ 
using the CFHTLenS and FastSound survey data. 
The GI cross-correlation function between galaxy positions and shapes yielded $2.4\sigma$ level signals for the red galaxies. 
The constraint on the amplitude IA for red galaxies,
$27.48_{-11.54}^{+11.53}$, was larger than the value extrapolated from the low-redshift measurements. 
Ongoing and future galaxy surveys will greatly improve the precision at such high redshifts,
providing more information on the evolution of the IA.
By performing the fisher-matrix analysis, we reconfirmed at $z\sim1.3$,
that future lensing surveys would require the mitigation of IAs to deliver their best performances
in giving accurate cosmological implications. 

We did not find any signal of GI for blue galaxies, consistent with earlier studies \citep{Mandelbaum:2011, Tonegawa:2018}, 
while upcoming galaxy surveys such as the Subaru Prime Focus Spectrograph \citep{Takada:2014} target blue galaxies at high redshifts. 
Recently, an interesting work of \citet{Shi:2021} presented a method to quantify IA of blue galaxies. 
Thus, not only red galaxies but blue galaxies can be good tracers of the tidal field. 
Our analysis provides the first step to utilize IA measurements in future surveys as a powerful probe of cosmology.

\acknowledgments

We thank Stephen Appleby for helpful comments on the manuscript.
The FastSound project was supported in part by MEXT/JSPS KAKENHI Grant Numbers 19740099, 19035005, 20040005, 22012005, and 23684007. 
This work is in part based on data collected at Subaru Telescope, which is operated by the National Astronomical Observatory of Japan.
This work is based on observations obtained with MegaPrime/MegaCam, a joint project of CFHT and CEA/IRFU, at the Canada-France-Hawaii Telescope (CFHT) which is operated by the National Research Council (NRC) of Canada, the Institut National des Sciences de l'Univers of the Centre National de la Recherche Scientifique (CNRS) of France, and the University of Hawaii. This research used the facilities of the Canadian Astronomy Data Centre operated by the National Research Council of Canada with the support of the Canadian Space Agency. CFHTLenS data processing was made possible thanks to significant computing support from the NSERC Research Tools and Instruments grant program.
MT is supported by an appointment to the JRG Program at the APCTP through the Science and Technology Promotion Fund and Lottery Fund of the Korean Government, and was also supported by the Korean Local Governments in Gyeongsangbuk-do Province and Pohang City.
TO acknowledges support from the Ministry of Science and Technology of Taiwan under Grants No. MOST 
110-2112-M-001-045-
and the Career Development Award, Academia Sinica (AS-CDA-108-M02) for the period of 2019 to 2023.
The FastSound project was supported in part by MEXT/JSPS KAKENHI Grant Numbers 19740099, 19035005, 20040005, 22012005, and 23684007. 
This work is in part based on data collected at Subaru Telescope, which is operated by the National Astronomical Observatory of Japan.






\appendix

\section{systematics tests}\label{section:appendix_systematics}

\begin{figure}[bt]\label{figure:systematics}
 \begin{center}
  \includegraphics[width=0.45\textwidth]{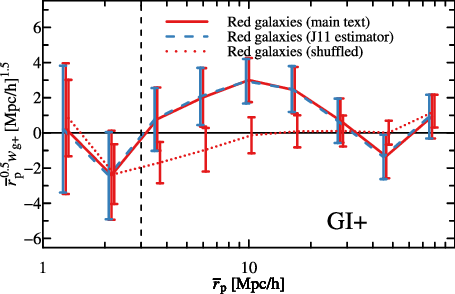}
 \end{center}
 \caption{Projected correlation functions $w_{g+}(\bar{r}_p)$ measured with different settings as a diagnosis for systematics. The solid red line is identical to that in figure \ref{figure:GI}. The dashed blue line is obtained by using the estimator \eqref{equation:estimator_M06}. The dotted red line is the same as the solid red line except that the ellipticity of the shape sample is shuffled.
}
\label{figure:systematics}
\end{figure}

In this appendix, we provide some tests to ensure that our estimate of the GI correlation is not affected by systematic effects.
First, we present the measurements of $w_{g+}$ using two estimators, equations \eqref{equation:estimator_H93} and \eqref{equation:estimator_M06}. 
The solid red and dashed blue lines of figure \ref{figure:systematics} compare $w_{g+}$ from the two estimators, 
respectively.
There is a negligibly small difference, confirming that our result is unchanged by choice of the estimators.

Another test involves the shuffling of ellipticity values in the shape sample.
After the randomization, the shape should no longer correlate with the density field, and $w_{g+}$ is expected to be zero. 
We show the result in figure \ref{figure:systematics} as the dotted red line. It is indeed consistent with zero at $\bar{r}_p>3 {\rm Mpc}/h$, where we assumed that our data are free from the systematics,
indicating that our observed GI signal manifests the true intrinsic shear-density correlation.

\section{Theoretical modeling of correlation functions}

In this appendix, following \citet{Joachimi:2011}, we summarize the modeling of the correlation function, including photo-$z$ uncertainties (Appendix \ref{sec:photoz}) and contributions of gravitational lensing under the Limber approximation (Appendix \ref{sec:limber}).

\subsection{The effect of photo-$z$ uncertainties}\label{sec:photoz}
If we denote quantities determined via photo-$z$ by a bar, the GI correlation function measured in a photo-$z$ survey is expressed by an integral of the true one, $\xi_{g+}^{\spec}$, as
\begin{equation}\label{equation:xigpphot}
\xi_{\rm gI}
 (\bar{r}_p,\bar{r}_\pi,\bar{z}_{\rm m}) = \int dz_{\rm m} \int dr_p \int dr_\pi
 p(r_p,r_\pi,z_m | \bar{r}_p,\bar{r}_\pi,\bar{z}_m) \xi_{\rm gI}^{\rm spec}(r_p,r_\pi,z_{\rm m}),
\end{equation}
where 
$p(r_p,r_\pi,z_m | \bar{r}_p,\bar{r}_\pi,\bar{z}_m)$ is the conditional probability distribution of $(r_p,r_\pi,z_{\rm m})$ for given $(\bar{r}_p,\bar{r}_\pi,\bar{z}_{\rm m})$
and
$\zm$ is the mean redshift of the density ($z_1$) and shape ($z_2$) samples, 
%\be
$
\zm = \frac{1}{2}(z_1+z_2)
$. 
We use the approximation for the pair separations $(r_p,r_\pi)$,
as 
$r_p \approx \theta \chi(z_{\rm m})$ and 
$r_\pi \approx (c/H(z_{\rm m}))(z_2-z_1)$,
and change the variables from $(r_p,r_\pi)$ to $(z_1,z_2)$.
We then obtain the expression, 
\begin{equation}\label{equation:xigpphot_final}
\xi_{\rm gI}(\bar{r}_p,\bar{r}_\pi,\bar{z}_{\rm m}) = 
\int dz_1 \int dz_2 p_n\left(z_1|\bar{z}_1\right) p_\epsilon\left(z_2|\bar{z}_2\right)
\xi_{\rm gI}^{\rm spec}\left(\bar{r}_p\frac{\chi\left(z_{\rm m}\right)}{\chi(\bar{z}_{\rm m})},\frac{c(z_2-z_1)}{H\left(z_{\rm m}\right)},z_{\rm m}\right),
\end{equation}
where $\zm$ is given 
above
and $\bar{z}_{1,2}= \bar{z}_{\rm m}\mp \bar{r}_\pi H(\bar{z}_{\rm m })/2c $. 
To obtain equation (\ref{equation:xigpphot_final}), we have assumed that photo-$z$ affects only the radial distances but not the angular positions.
The factor $\chi(\zm))/\chi(\bar{z}_{\rm m})$ arises from the assumption that $\theta$ is conserved.
In our analysis, the density sample has spectroscopic redshifts. 
Thus setting $p_n(z_1|\bar{z}_1)=\delta^{\rm D}(z_1-\bar{z}_1)$ leads to equation \eqref{equation:xig+bar}.

\subsection{Limber Approximations}\label{sec:limber}

The cross-correlation of galaxy position and ellipticity, denoted as the $n\epsilon$ term, contains contributions not only from the gI correlation but also from galaxy-galaxy lensing (gG), magnification-shear correlations (mG), and magnification-intrinsic ellipticity correlations (mI). In terms of projected angular power spectra including the photo-$z$'s, $\bar{z}_1$ and $\bar{z}_2$, we can write it as 
\be
C_{n\epsilon}(\ell;\bar{z}_1,\bar{z}_2) = 
C_{\rm gI}(\ell;\bar{z}_1,\bar{z}_2) + 
C_{\rm gG}(\ell;\bar{z}_1,\bar{z}_2) + 
C_{\rm mG}(\ell;\bar{z}_1,\bar{z}_2) + 
C_{\rm mI}(\ell;\bar{z}_1,\bar{z}_2). \label{eq:angular_ne}
\ee
Under the Limber approximations, these angular power spectra are given by \citep[e.g.,][]{Joachimi:2010},
\begin{align}\label{eq:limber_2d}
    C_{\rm gI}(\ell;\bar{z}_1,\bar{z}_2) &= b_g\int^{\chi_{\rm hor}}_0 d\chi' \frac{p_n(\chi'|\chi(\bar{z}_1)) p_\epsilon(\chi'|\chi(\bar{z}_2))}{\chi^{\prime 2}} P_{\delta {\rm I}}\left( \frac{\ell}{\chi'},z(\chi')\right), \\
    C_{\rm gG}(\ell;\bar{z}_1,\bar{z}_2) &= b_g\int^{\chi_{\rm hor}}_0 d\chi' \frac{p_n(\chi'|\chi(\bar{z}_1)) q_\epsilon(\chi',\chi(\bar{z}_2))}{\chi^{\prime 2}} P_{\delta}\left( \frac{\ell}{\chi'},z(\chi')\right), \nn \\    
    C_{\rm mG}(\ell;\bar{z}_1,\bar{z}_2) &= 2(\alpha_s-1)\int^{\chi_{\rm hor}}_0 d\chi' \frac{q_n(\chi',\chi(\bar{z}_1)) q_\epsilon(\chi',\chi(\bar{z}_2))}{\chi^{\prime 2}} P_{\delta}\left( \frac{\ell}{\chi'},z(\chi')\right),\nn \\    
    C_{\rm mI}(\ell;\bar{z}_1,\bar{z}_2) &= 2(\alpha_s-1)\int^{\chi_{\rm hor}}_0 d\chi' \frac{q_n(\chi',\chi(\bar{z}_1)) p_\epsilon(\chi'|\chi(\bar{z}_2))}{\chi^{\prime 2}} P_{\delta {\rm I}}\left( \frac{\ell}{\chi'},z(\chi')\right),\\
\end{align}
where $\chi_{\rm hor}$ is the comoving horizon distance, $\alpha_s$ is the logarithmic slope of the cumulative galaxy luminosity function of the density sample, and $q_x$ ($x=\{n,\epsilon\}$) is the lensing weight function, 
\be
q_x(\chi,\chi_1) = \frac{3H_0^2\Omega_m}{2c^2}\frac{\chi}{a(\chi)}
\int^{\chi_{\rm hor}}_\chi d\chi' p_x(\chi'|\chi_1)\frac{\chi'-\chi}{\chi'}.
\ee
Using these angular power spectra, the three-dimensional correlation functions are obtained as 
\begin{align}
    \xi_X (\bar{r}_p, \bar{r}_\pi;\bar{z}_{\rm m}) = -\int^\infty_0 \frac{d\ell}{2\pi}\ell J_2(\ell\theta(\bar{r}_p,\bar{z}_{\rm m})) 
    C_X(\ell; \bar{z}_1(\bar{z}_{\rm m},\bar{r}_\pi),\bar{z}_2(\bar{z}_{\rm m},\bar{r}_\pi)),
\end{align}
where $X=\{ {\rm gI}, {\rm gG}, {\rm mG},{\rm mI} \}$. 
By comparing $\xi_{\rm gI}$ obtained in this way to the full expression, equation~\eqref{equation:xigpphot_final}, we have verified that the Limber approximation is accurate enough within the measurement uncertainties.

\bibliographystyle{apj}
\bibliography{refs}

\begin{thebibliography}{}
\expandafter\ifx\csname natexlab\endcsname\relax\def\natexlab#1{#1}\fi

\bibitem[{{Ben{\'\i}tez}(2000)}]{Benitez:2000}
{Ben{\'\i}tez}, N. 2000, \apj, 536, 571

\bibitem[{{Blazek} {et~al.}(2011){Blazek}, {McQuinn}, \&
  {Seljak}}]{Blazek:2011}
{Blazek}, J., {McQuinn}, M., \& {Seljak}, U. 2011, \jcap, 5, 10

\bibitem[{{Bridle} \& {King}(2007)}]{Bridle:2007}
{Bridle}, S., \& {King}, L. 2007, New Journal of Physics, 9, 444

\bibitem[{{Capak}(2004)}]{Capak:2004}
{Capak}, P.~L. 2004, PhD thesis, UNIVERSITY OF HAWAI'I

\bibitem[{{Catelan} {et~al.}(2001){Catelan}, {Kamionkowski}, \&
  {Blandford}}]{Catelan:2001}
{Catelan}, P., {Kamionkowski}, M., \& {Blandford}, R.~D. 2001, \mnras, 320, L7

\bibitem[{{Chisari} {et~al.}(2016){Chisari}, {Laigle}, {Codis}, {Dubois},
  {Devriendt}, {Miller}, {Benabed}, {Slyz}, {Gavazzi}, \&
  {Pichon}}]{Chisari:2016a}
{Chisari}, N., {Laigle}, C., {Codis}, S., {et~al.} 2016, \mnras, 461, 2702

\bibitem[{{Chisari} \& {Dvorkin}(2013)}]{Chisari:2013}
{Chisari}, N.~E., \& {Dvorkin}, C. 2013, \jcap, 12, 029

\bibitem[{{Coleman} {et~al.}(1980){Coleman}, {Wu}, \& {Weedman}}]{Coleman:1980}
{Coleman}, G.~D., {Wu}, C.~C., \& {Weedman}, D.~W. 1980, \apjs, 43, 393

\bibitem[{{Croft} \& {Metzler}(2000)}]{Croft:2000}
{Croft}, R.~A.~C., \& {Metzler}, C.~A. 2000, \apj, 545, 561

\bibitem[{{Du} {et~al.}(2020){Du}, {Cheng}, {Zheng}, \& {Wu}}]{Du:2020}
{Du}, W., {Cheng}, C., {Zheng}, Z., \& {Wu}, H. 2020, \aj, 159, 138

\bibitem[{{Erben} {et~al.}(2013){Erben}, {Hildebrandt}, {Miller}, {van
  Waerbeke}, {Heymans}, {Hoekstra}, {Kitching}, {Mellier}, {Benjamin}, {Blake},
  {Bonnett}, {Cordes}, {Coupon}, {Fu}, {Gavazzi}, {Gillis}, {Grocutt}, {Gwyn},
  {Holhjem}, {Hudson}, {Kilbinger}, {Kuijken}, {Milkeraitis}, {Rowe},
  {Schrabback}, {Semboloni}, {Simon}, {Smit}, {Toader}, {Vafaei}, {van Uitert},
  \& {Velander}}]{Erben:2013}
{Erben}, T., {Hildebrandt}, H., {Miller}, L., {et~al.} 2013, \mnras, 433, 2545

\bibitem[{{Fioc} \& {Rocca-Volmerange}(1999)}]{Fioc:1999}
{Fioc}, M., \& {Rocca-Volmerange}, B. 1999, arXiv e-prints,
  arXiv:astro-ph/9912179

\bibitem[{{Hamilton}(1993)}]{Hamilton:1993a}
{Hamilton}, A.~J.~S. 1993, \apj, 417, 19

\bibitem[{{Hamilton}(1998)}]{Hamilton:1998}
---. 1998, Astrophysics and Space Science Library, Vol. 231, {Linear Redshift
  Distortions: a Review}, ed. D.~{Hamilton}, 185

\bibitem[{{Heavens} {et~al.}(2000){Heavens}, {Refregier}, \&
  {Heymans}}]{Heavens:2000}
{Heavens}, A., {Refregier}, A., \& {Heymans}, C. 2000, \mnras, 319, 649

\bibitem[{{Hildebrandt} {et~al.}(2012){Hildebrandt}, {Erben}, {Kuijken}, {van
  Waerbeke}, {Heymans}, {Coupon}, {Benjamin}, {Bonnett}, {Fu}, {Hoekstra},
  {Kitching}, {Mellier}, {Miller}, {Velander}, {Hudson}, {Rowe}, {Schrabback},
  {Semboloni}, \& {Ben{\'\i}tez}}]{Hildebrandt:2012}
{Hildebrandt}, H., {Erben}, T., {Kuijken}, K., {et~al.} 2012, \mnras, 421, 2355

\bibitem[{{Hirata} {et~al.}(2007){Hirata}, {Mandelbaum}, {Ishak}, {Seljak},
  {Nichol}, {Pimbblet}, {Ross}, \& {Wake}}]{Hirata:2007}
{Hirata}, C.~M., {Mandelbaum}, R., {Ishak}, M., {et~al.} 2007, \mnras, 381,
  1197

\bibitem[{{Hirata} \& {Seljak}(2004)}]{Hirata:2004}
{Hirata}, C.~M., \& {Seljak}, U. 2004, \prd, 70, 063526

\bibitem[{{Hirata} \& {Seljak}(2010)}]{Hirata:2010}
---. 2010, \prd, 82, 049901

\bibitem[{{Iwamuro} {et~al.}(2012){Iwamuro}, {Moritani}, {Yabe}, {Sumiyoshi},
  {Kawate}, {Tamura}, {Akiyama}, {Kimura}, {Takato}, {Tait}, {Ohta}, {Totani},
  {Suzuki}, \& {Tonegawa}}]{Iwamuro:2012}
{Iwamuro}, F., {Moritani}, Y., {Yabe}, K., {et~al.} 2012, \pasj, 64, 59

\bibitem[{{Joachimi} \& {Bridle}(2010)}]{Joachimi:2010}
{Joachimi}, B., \& {Bridle}, S.~L. 2010, \aap, 523, A1

\bibitem[{{Joachimi} {et~al.}(2011){Joachimi}, {Mandelbaum}, {Abdalla}, \&
  {Bridle}}]{Joachimi:2011}
{Joachimi}, B., {Mandelbaum}, R., {Abdalla}, F.~B., \& {Bridle}, S.~L. 2011,
  \aap, 527, A26

\bibitem[{{Joachimi} {et~al.}(2015){Joachimi}, {Cacciato}, {Kitching},
  {Leonard}, {Mandelbaum}, {Sch{\"a}fer}, {Sif{\'o}n}, {Hoekstra}, {Kiessling},
  {Kirk}, \& {Rassat}}]{Joachimi:2015}
{Joachimi}, B., {Cacciato}, M., {Kitching}, T.~D., {et~al.} 2015, \ssr, 193, 1

\bibitem[{{Johnston} {et~al.}(2019){Johnston}, {Georgiou}, {Joachimi},
  {Hoekstra}, {Chisari}, {Farrow}, {Fortuna}, {Heymans}, {Joudaki}, {Kuijken},
  \& {Wright}}]{Johnston:2019}
{Johnston}, H., {Georgiou}, C., {Joachimi}, B., {et~al.} 2019, \aap, 624, A30

\bibitem[{{Johnston} {et~al.}(2021){Johnston}, {Joachimi}, {Norberg},
  {Hoekstra}, {Eriksen}, {Fortuna}, {Manzoni}, {Serrano}, {Siudek},
  {Tortorelli}, {Asorey}, {Cabayol}, {Carretero}, {Casas}, {Castander},
  {Crocce}, {Fernandez}, {Garc{\'\i}a-Bellido}, {Gaztanaga}, {Hildebrandt},
  {Miquel}, {Navarro-Girones}, {Padilla}, {Sanchez}, {Sevilla-Noarbe}, \&
  {Tallada-Cresp{\'\i}}}]{Johnston:2021}
{Johnston}, H., {Joachimi}, B., {Norberg}, P., {et~al.} 2021, \aap, 646, A147

\bibitem[{{Kirk} {et~al.}(2015){Kirk}, {Brown}, {Hoekstra}, {Joachimi},
  {Kitching}, {Mandelbaum}, {Sif{\'o}n}, {Cacciato}, {Choi}, {Kiessling},
  {Leonard}, {Rassat}, \& {Sch{\"a}fer}}]{Kirk:2015}
{Kirk}, D., {Brown}, M.~L., {Hoekstra}, H., {et~al.} 2015, \ssr, 193, 139

\bibitem[{{Kurita} {et~al.}(2021){Kurita}, {Takada}, {Nishimichi}, {Takahashi},
  {Osato}, \& {Kobayashi}}]{Kurita:2021}
{Kurita}, T., {Takada}, M., {Nishimichi}, T., {et~al.} 2021, \mnras, 501, 833

\bibitem[{{Mahajan} {et~al.}(2018){Mahajan}, {Drinkwater}, {Driver}, {Hopkins},
  {Graham}, {Brough}, {Brown}, {Holwerda}, {Owers}, \&
  {Pimbblet}}]{Mahajan:2018}
{Mahajan}, S., {Drinkwater}, M.~J., {Driver}, S., {et~al.} 2018, \mnras, 475,
  788

\bibitem[{{Mandelbaum} {et~al.}(2006){Mandelbaum}, {Hirata}, {Ishak}, {Seljak},
  \& {Brinkmann}}]{Mandelbaum:2006}
{Mandelbaum}, R., {Hirata}, C.~M., {Ishak}, M., {Seljak}, U., \& {Brinkmann},
  J. 2006, \mnras, 367, 611

\bibitem[{{Mandelbaum} {et~al.}(2011){Mandelbaum}, {Blake}, {Bridle},
  {Abdalla}, {Brough}, {Colless}, {Couch}, {Croom}, {Davis}, {Drinkwater},
  {Forster}, {Glazebrook}, {Jelliffe}, {Jurek}, {Li}, {Madore}, {Martin},
  {Pimbblet}, {Poole}, {Pracy}, {Sharp}, {Wisnioski}, {Woods}, \&
  {Wyder}}]{Mandelbaum:2011}
{Mandelbaum}, R., {Blake}, C., {Bridle}, S., {et~al.} 2011, \mnras, 410, 844

\bibitem[{{Miller} {et~al.}(2007){Miller}, {Kitching}, {Heymans}, {Heavens}, \&
  {van Waerbeke}}]{Miller:2007}
{Miller}, L., {Kitching}, T.~D., {Heymans}, C., {Heavens}, A.~F., \& {van
  Waerbeke}, L. 2007, \mnras, 382, 315

\bibitem[{{Okada} {et~al.}(2016){Okada}, {Totani}, {Tonegawa}, {Akiyama},
  {Dalton}, {Glazebrook}, {Iwamuro}, {Ohta}, {Takato}, {Tamura}, {Yabe},
  {Bunker}, {Goto}, {Hikage}, {Ishikawa}, {Okumura}, \& {Shimizu}}]{Okada:2016}
{Okada}, H., {Totani}, T., {Tonegawa}, M., {et~al.} 2016, \pasj, 68, 47

\bibitem[{{Okumura} {et~al.}(2021){Okumura}, {Hayashi}, {Chiu}, {Lin}, {Osato},
  {Hsieh}, \& {Lin}}]{Okumura:2021}
{Okumura}, T., {Hayashi}, M., {Chiu}, I.~N., {et~al.} 2021, \pasj, 73, 1186

\bibitem[{{Okumura} \& {Jing}(2009)}]{Okumura:2009a}
{Okumura}, T., \& {Jing}, Y.~P. 2009, \apjl, 694, L83

\bibitem[{{Okumura} {et~al.}(2009){Okumura}, {Jing}, \& {Li}}]{Okumura:2009}
{Okumura}, T., {Jing}, Y.~P., \& {Li}, C. 2009, \apj, 694, 214

\bibitem[{{Okumura} \& {Taruya}(2020)}]{Okumura:2020}
{Okumura}, T., \& {Taruya}, A. 2020, \mnras, 493, L124

\bibitem[{{Okumura} {et~al.}(2020){Okumura}, {Taruya}, \&
  {Nishimichi}}]{Okumura:2020a}
{Okumura}, T., {Taruya}, A., \& {Nishimichi}, T. 2020, \mnras, 494, 694

\bibitem[{{Okumura} {et~al.}(2016){Okumura}, {Hikage}, {Totani}, {Tonegawa},
  {Okada}, {Glazebrook}, {Blake}, {Ferreira}, {More}, {Taruya}, {Tsujikawa},
  {Akiyama}, {Dalton}, {Goto}, {Ishikawa}, {Iwamuro}, {Matsubara},
  {Nishimichi}, {Ohta}, {Shimizu}, {Takahashi}, {Takato}, {Tamura}, {Yabe}, \&
  {Yoshida}}]{Okumura:2016}
{Okumura}, T., {Hikage}, C., {Totani}, T., {et~al.} 2016, \pasj, 68, 38

\bibitem[{{Samuroff} {et~al.}(2021){Samuroff}, {Mandelbaum}, \&
  {Blazek}}]{Samuroff:2021}
{Samuroff}, S., {Mandelbaum}, R., \& {Blazek}, J. 2021, \mnras, 508, 637

\bibitem[{{Scalo}(1986)}]{Scalo:1986}
{Scalo}, J.~M. 1986, \fcp, 11, 1

\bibitem[{{Schlegel} {et~al.}(1998){Schlegel}, {Finkbeiner}, \&
  {Davis}}]{Schlegel:1998}
{Schlegel}, D.~J., {Finkbeiner}, D.~P., \& {Davis}, M. 1998, \apj, 500, 525

\bibitem[{{Shi} {et~al.}(2021){Shi}, {Osato}, {Kurita}, \& {Takada}}]{Shi:2021}
{Shi}, J., {Osato}, K., {Kurita}, T., \& {Takada}, M. 2021, \apj, 917, 109

\bibitem[{{Singh} {et~al.}(2015){Singh}, {Mandelbaum}, \& {More}}]{Singh:2015}
{Singh}, S., {Mandelbaum}, R., \& {More}, S. 2015, \mnras, 450, 2195

\bibitem[{{Takada} {et~al.}(2014){Takada}, {Ellis}, {Chiba}, {Greene},
  {Aihara}, {Arimoto}, {Bundy}, {Cohen}, {Dor{\'e}}, {Graves}, {Gunn},
  {Heckman}, {Hirata}, {Ho}, {Kneib}, {F{\`e}vre}, {Lin}, {More}, {Murayama},
  {Nagao}, {Ouchi}, {Seiffert}, {Silverman}, {Sodr{\'e}}, {Spergel}, {Strauss},
  {Sugai}, {Suto}, {Takami}, \& {Wyse}}]{Takada:2014}
{Takada}, M., {Ellis}, R.~S., {Chiba}, M., {et~al.} 2014, \pasj, 66, 1

\bibitem[{{Takahashi} {et~al.}(2012){Takahashi}, {Sato}, {Nishimichi},
  {Taruya}, \& {Oguri}}]{Takahashi:2012}
{Takahashi}, R., {Sato}, M., {Nishimichi}, T., {Taruya}, A., \& {Oguri}, M.
  2012, \apj, 761, 152

\bibitem[{{Taruya} \& {Okumura}(2020)}]{Taruya:2020}
{Taruya}, A., \& {Okumura}, T. 2020, \apjl, 891, L42

\bibitem[{{Tonegawa} {et~al.}(2018){Tonegawa}, {Okumura}, {Totani}, {Dalton},
  {Glazebrook}, \& {Yabe}}]{Tonegawa:2018}
{Tonegawa}, M., {Okumura}, T., {Totani}, T., {et~al.} 2018, \pasj, 70, 41

\bibitem[{{Tonegawa} {et~al.}(2015){Tonegawa}, {Totani}, {Okada}, {Akiyama},
  {Dalton}, {Glazebrook}, {Iwamuro}, {Maihara}, {Ohta}, {Shimizu}, {Takato},
  {Tamura}, {Yabe}, {Bunker}, {Coupon}, {Ferreira}, {Frenk}, {Goto}, {Hikage},
  {Ishikawa}, {Matsubara}, {More}, {Okumura}, {Percival}, {Spitler}, \&
  {Szapudi}}]{Tonegawa:2015}
{Tonegawa}, M., {Totani}, T., {Okada}, H., {et~al.} 2015, \pasj, 67, 81

\bibitem[{{Wang} {et~al.}(2011){Wang}, {Jing}, {Li}, {Okumura}, \&
  {Han}}]{Wang:2011}
{Wang}, W., {Jing}, Y.~P., {Li}, C., {Okumura}, T., \& {Han}, J. 2011, \apj,
  734, 88

\bibitem[{{Yao} {et~al.}(2020){Yao}, {Shan}, {Zhang}, {Kneib}, \&
  {Jullo}}]{Yao:2020}
{Yao}, J., {Shan}, H., {Zhang}, P., {Kneib}, J.-P., \& {Jullo}, E. 2020, \apj,
  904, 135

\end{thebibliography}

\end{document}